\documentstyle[psfig,conf_iap,10pt]{article}
\def\lhf{\hbox{$\scriptstyle\rm HI$}}
\def\B{\hbox{$\scriptstyle\rm B$}}
\def\Q{\hbox{$\scriptstyle\rm Q$}}
\def\QB{\hbox{$\scriptstyle\rm Q/\rm B\ $}}
\newcommand{\cmm}{\,{\rm cm^{-2}}}
\newcommand{\ltsima}{$\; \buildrel < \over \sim \;$}
\def\lsim{\lower.5ex\hbox{\ltsima}}

\begin{document}
\heading{%
Proximity Effect for Metal Absorption Systems}
\par\medskip\noindent
\author{%
J.M.Liu$^{1}$}
\address{%
SISSA, Via Beirut n.2-4, 34014 Trieste, Italy}
\begin{abstract}
Using our photoionization code CIRRUS \cite{LiuM97}, we show that 
the study of proximity effect on metal elements provides
a possibility of placing constraints on the spectral shape of the 
metagalactic UV background (MUVB).
From theoretical point of view, the most optimum indicator of
the spectral shape of the MUVB is the ratio O~{\sc iv}/O~{\sc iii}.
Unlike previous thought, the ratio Si~{\sc iv}/C~{\sc iv} is not a good 
indicator of the spectral shape because of its strong dependence 
on an ionization parameter as well. We also find that the observed excess 
of C~{\sc iv} systems in luminous QSOs may be a photoionization effect,
but not a gravitational lensing effect \cite{VanQ96}.
\end{abstract}
\section{Introduction}
The H~{\sc i} proximity effect has been widely studied since it was 
discovered. One thing of particular 
interest is that the study of the H~{\sc i} proximity
effect provides a method of measuring the intensity of the MUVB
at the H~{\sc i} edge. However, the H~{\sc i} proximity effect
does not reveal information on the spectral shape of the MUVB.\par

We demonstrate that the proximity effect 
on metal elements may provide a method for constraining 
the spectral shape of the MUVB. Songalia and Cowie (1996) 
\cite{SonC96} reported that 
metal elements are detected in 75\% of clouds with $N_{\lhf} >3.0\times 10^{14} 
\cmm$. The lack of data for studying the proximity effect 
on metal elements is going to overcome. If future observations
detect what we expect for the proximity effect on metal elements,
it will confirm that the proximity effect is an ionization effect
and photoionization is the dominant ionization mechanism for metal
ions in which the proximity effect is seen. 
\section{Results}
The electronic configurations of metal elements are complicated
and a large number of ionization stages can coexist simultaneously.
In general, a simple feature does not exist for metal proximity effect, 
unlike for the H~{\sc i} proximity effect. Therefore, we may study the 
dependence of the column density ratios of two ions in successive 
ionization stages on $\omega$.\par
We found that [O~{\sc iv}/O~{\sc iii}]$_{\QB}$ is the most optimum 
indicator of the spectral shape of the MUVB (we use [X]$_{\QB}$ to represent 
the ratio of the value of X in the vicinity of the QSO to that
in background case). The HST does provide
the possibility to observe O~{\sc iv} and O~{\sc iii} simultaneously 
\cite{TriL97} and \cite{VogR95}.
It is clear from Figure 1 that [O~{\sc iv}/O~{\sc iii}]$_{\QB}$
strongly depends on the assumed spectral shape of the MUVB
and less sensitive to the assumed $U_{\B}$ (the ionization parameter
in background case) 
in small $\omega$ region. Therefore, [O~{\sc iv}/O~{\sc iii}]$_{\QB}$ 
is good for constraining the spectral shape even if $U_{\B}$ is unknown.
\begin{figure}
\centerline{\vbox{
\psfig{figure=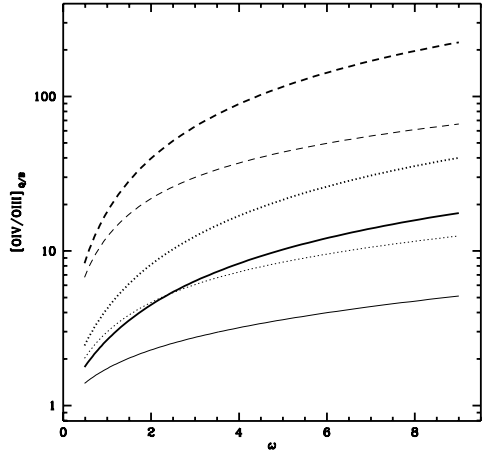,height=7.cm}
}}
\caption[]{[O~{\sc iv}/O~{\sc iii}]$_{\QB}$ vs $\omega$. 
The heavy curves are for $U_{\B}=10^{-2.7}$; the light
curves are for $U_{\B}=10^{-1.5}$.
The solid curves are for $\alpha_{\B}=\alpha_{\Q}=1.5$ and $b=1$; 
the dotted curves are for $\alpha_{\B}=0.5$,
$b=12.5$, and $\alpha_{\Q}=1.5$; the dashed curves are for
$\alpha_{\B}=0.3$, $b=100$, and $\alpha_{\Q}=1.5$.
}
\end{figure}
The ratio Si~{\sc iv}/C~{\sc iv}, contrast to previous thought, is
not a good indicator to the spectral shape, because it
strongly depends on both the ionization parameter $U_{\B}$ and
the spectral shape of the MUVB. A small uncertainty on either of them
must results in a big error in determination of the other.
We also found that the extra ionizing radiation from QSOs does not 
always reduce the ratio C~{\sc iv}/H~{\sc i}, but enhances 
C~{\sc iv}/H~{\sc i} in the case of $U_{\B}\lsim 10^{-2}$.
So, the observed excess of C~{\sc iv} systems in luminous
QSOs may be a photoionization effect,
but not a gravitational lensing effect \cite{VanQ96}.

\acknowledgements{We are grateful to Dr Avery Meiksin for his comments.}

\vfill

\begin{iapbib}{99}{
\bibitem{LiuM97} Liu, J.M., \& Meiksin, A., 1997, in preparation
\bibitem{SonC96} Songaila, A., \& Cowie, L.L. 1996, AJ, 112, 335
\bibitem{TriL97} Tripp, T.M., Lu, L., \& Savage, B.D. 1997, 
	[astro-ph/9703080]
\bibitem{VanQ96} Vanden Berk, D.E., Quashnock, J.M., York, D.G.,
	\& Yanny, B. 1996, 469, 78 
\bibitem{VogR95} Vogel, S., \& Reimers, D. 1995, A\&A, 294, 377				
}
\end{iapbib}
\end{document}